\documentclass[aps,prl,twocolumn,showpacs,subfigure,superscriptaddress,nobibnotes,nofootinbib]{revtex4-1}
\usepackage[dvips]{graphicx}
\usepackage{epsfig}
\usepackage{bm}   
\usepackage{dcolumn}
\usepackage{graphicx}
\usepackage{rotate}
\usepackage{amsfonts}
\usepackage[table]{xcolor}
\usepackage{tabularx}
\usepackage{footmisc,booktabs,amssymb}

\begin{document}

\title{Reply to Comment on ``New formulas for the $(-2)$ moment of the photo-absorption cross section, $\sigma_{_{-2}}$''}

\author{J. N.~Orce}
\email{coulex@gmail.com} 
\affiliation{Department of Physics and Astronomy, University of the Western Cape, P/B X17, Bellville, ZA-7535 South Africa}

\date{\today}

\maketitle

In the Comment to my paper~\cite{orce1}, von Neumann-Cosel suggests that the low-energy contribution to $\sigma_{_{-2}}$ caused by the pygmy 
dipole resonance (PDR) should provide a systematic upward correction of the order of 5-10\%. The author additionally 
performs a free fit to $\sigma_{_{-2}}$ data from polarizability ($^{120}$Sn~\cite{120Sn} and $^{208}$Pb~\cite{tamii}), 
photoneutron cross section ($^{68}$Ni~\cite{68ni1,68ni2}), and total nuclear photoabsorption 
($^{12}$C, $^{16}$O, $^{27}$Al and $^{40}$Ca~\cite{Ahrens1975}) studies. 
Different volume, $S_v$,  and surface-to-volume, $S_s/S_v$, coefficients of the symmetry energy 
to the ones given by Tian and collaborators~\cite{tian} are extracted. 
According to von Neumann-Cosel, the coefficients from the free fit ``may be better suited''. 

\vspace{0.1cm}
I  agree with the author that, because of the low-energy PDR contribution, a systematic upward correction to 
$\sigma_{_{-2}}$ can be expected 
in nuclei with neutron excess considering the $E_{\gamma}^{-2}$ weighting of $\sigma_{_{-2}}$. 
However, it is premature to claim a general PDR contribution to $\sigma_{_{-2}}$  based on 
four measurements only (inelastic proton scattering at relativistic energies 
of $^{120}$Sn~\cite{120Sn}, $^{208}$Pb~\cite{tamii} and  $^{90}$Zr~\cite{90Zr}, and 
a selected photon scattering measurement in $^{138}$Ba~\cite{138Ba}). Moreover, the author provides estimates for only two 
measurements ($^{120}$Sn and $^{208}$Pb). 
Additional measurements of the PDR contribution for a broader range of nuclei with neutron excess 
are clearly needed to deduce a systematic effect.

\vspace{0.1cm}
Furthermore, I see several arguments against the symmetry energy parameters extracted  by von Neumann-Cosel.

\vspace{0.1cm}
1) The author uses the nuclear photoabsorption data from five nuclei ($^{12}$C, $^{16}$O, $^{27}$Al, 
$^{40}$Ca~\cite{Ahrens1975,Ahrens1976} and $^{68}$Ni~\cite{68ni1,68ni2}) to perform a free fit to the $S_v$ and  
$S_s/S_v$ variables in Eq.[12] of my paper~\cite{orce1}. 
He claims that his resulting parameters ``may be better suited'' to the $\sigma_{_{-2}}$ trend partly because 
four data points ($^{12}$C, $^{16}$O, $^{27}$Al, 
$^{40}$Ca~\cite{Ahrens1975,Ahrens1976}) include both 
photoproton and photoneutron cross sections above 10 MeV. 
These four data points are derived from a single total nuclear photoabsorption study by Ahrens and co-workers in 1975, 
which used bremsstrahlung photon beams. 
As pointed out by Bergere~\cite{lectures59}, care must be taken concerning this method, because it has 
the drawback of large non-nuclear contributions (e.g., Compton scattering, pair production, dead times) which are several 
tens of times larger than the total nuclear photoabsorption cross section~\cite{lectures59}. 
Monte-Carlo simulations should be conducted to calculate the error for each non-nuclear contribution. 
Such simulations are not evident in Ref.~\cite{Ahrens1975}. 
Therefore, the less than 0.1\% error from non-nuclear effects claimed by Ahrens and collaborators is questionable. 
Moreover, if these measurements were as powerful and precise, one 
can only wonder why they were not verified and applied for the photon energy range of interest to $\sigma_{_{-2}}$ since. 

In addition, the $^{68}$Ni data point~\cite{68ni1,68ni2} in Fig. 1 of the Comment only includes $(\gamma,n)$ and $(\gamma,2n)$ photoneutron 
cross sections and does not account for other neutron decay channels and photoproton contributions. 
More relevantly, most of the existing information on photoabsorption cross sections arise from stable nuclei, i.e., 
we know very little on how unbound nuclei polarize. Hence, the $^{68}$Ni data point should not be included in the free fit.

\begin{figure*}[]
\begin{center}
\includegraphics[width=11.2cm,height=8cm,angle=-0]{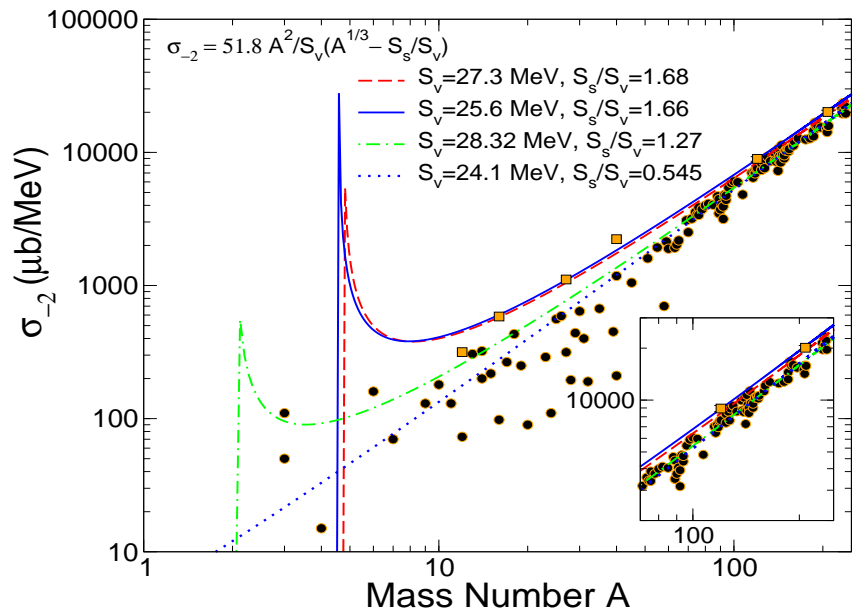} 
\caption{(Color online) $\sigma_{_{-2}}$ 
plots for different sets of ($S_v, S_s/S_v$) symmetry-energy parameters discussed in this work. 
The $\sigma_{_{-2}}$ data $vs$ A  on a log-log scale from the 1988 photo-neutron 
cross-section evaluation (circles)~\cite{atlas} and from data provided in the Comment (squares)
are shown for comparison. From the data presented in Fig. [1] of the Comment, the $^{68}$Ni data point has been 
removed (see text for explanation) and the $^{40}$Ca data point has not been lowered from Ref.~\cite{Ahrens1975}.}
\label{fig:sets}
\end{center}
\end{figure*}

\vspace{0.1cm}
2) In Ref.~\cite{orce1}, I used two independent methods to derive $\sigma_{_{-2}}$: 1) from a fit to the extensive 
photoneutron compilation published in 1988~\cite{atlas}, which includes data from the preferred method of monochromatic photon 
beams generated by in-flight annihilation of positrons\footnote{One main advantage of this technique over bremsstrahlung 
photon beams is the direct and simultaneous measurements of the partial photoneutron cross sections which are in competition 
in the {\small GDR} region.}, and shows overall agreement and consistency between 
measurements done at Livermore, Giessen, Saclay and other laboratories, and 2) from the mass dependence of the 
symmetry energy extracted from a global fit to the binding energies of 
isobaric nuclei with $A\geq10$~\cite{tian} given by the 2012 mass evaluation~\cite{audi}. 
These two  predicted trends smoothly converge with the $\sigma_{_{-2}}$ data~\cite{atlas} above $A\gtrsim70$, 
in agreement with the dominant photoneutron cross sections for heavy nuclei. 
No consistency in the photoneutron data is observed for lighter nuclei, which highlights the necessity for 
systematic studies of photoproton cross sections for $A\lesssim70$ nuclei. 
This should, preferably, be done in direct and simultaneous measurements of the partial photoneutron 
and photoproton cross sections, crucial to obtain reliable total photonuclear cross sections, as described in Ref.~\cite{lectures75}.




3) Figure~\ref{fig:sets} shows $\sigma_{_{-2}}$ 
plots for the different sets of symmetry energy parameters discussed in this work. 
Von Neumann-Cosel   shows that, when omitting the $^{12}$C data point, his analysis provides a better fit to the available data with 
symmetry-energy parameters ($S_v= 25.6(8)$ MeV, $S_s/S_v = 1.66(5)$) similar to those calculated in Ref.~\cite{steiner} 
($S_v= 27.3$ MeV, $S_s/S_v=1.68$). 
One should at least mention the drawback of these theoretical parameters 
which include the Coulomb interaction of protons but do not imply a neutron skin, 
later precisely measured in $^{208}$Pb by Tamii and collaborators~\cite{tamii}. 
Considering a neutron skin has a dramatic effect on the calculated surface-to-volume ratio ($S_v=24.1$ MeV, $S_s/S_v=0.545$)~\cite{steiner}.  
It is true that the latter  parameters fail to describe the $\sigma_{_{-2}}$ data 
for light nuclei, but it seems to work where it is intended to, i.e., for heavy nuclei, where the 
excess neutrons can form a skin against a $N\approx Z$ core. In fact, 
the calculated $\sigma_{_{-2}}$ trend implying a neutron skin ($S_v=24.1$ MeV, $S_s/S_v=0.545$) also converges with 
the photoneutron data~\cite{atlas} and with Eq. [14]  ($S_v= 28.32$ MeV, $S_s/S_v = 1.27$) in Ref.~\cite{orce1} 
for $A\gtrsim70$, as clearly shown  in the inset of Fig.~\ref{fig:sets}.



%
%
%
%
%

In conclusion, I  agree with the relevance of the PDR contribution at low energies~\cite{atlas}, a contribution that 
remains to be quantified for many heavy nuclei with neutron excess, but 
disagree with the statement that the different volume and surface-to-volume coefficients of the symmetry energy 
extracted from a free fit in von Neunman-Cosel's Comment are better suited than the ones 
chosen in my paper~\cite{tian}. The fact of the matter 
is that additional data  are vital to pin down the mass dependence of $\sigma_{_{-2}}$ and the symmetry energy, especially for nuclei 
below $A\approx70$. The author acknowledges funding support by the South African National 
Research Foundation (NRF) under Grant 93500.

\end{document}